\documentclass[preprint]{revtex4}
\usepackage{graphicx}
\newcounter{subfigure}
\begin{document}

\begin{titlepage}
\title{Glassy Aging with modified Kohlrausch-Williams-Watts form}.\\
\author{Bhaskar Sen Gupta and Shankar P. Das}

\affiliation{School of Physical Sciences,
Jawaharlal Nehru University,\\
New Delhi 110067, India.}

\vspace*{1.5cm}
\begin{abstract}
In this report we address the question whether aging in the non
equilibrium glassy state is controlled by the
equilibrium $\alpha$-relaxation process which occur at temperatures above
$T_g$. Recently
Lunkenheimer {\em et. al.} [Phys. Rev.
Lett. {\bf 95}, 055702 (2005)] proposed a model for  the
glassy aging data of dielectric relaxation using a modified
Kohlrausch-Williams-Watts (KWW) form $\exp\left[-{(t_{\mathrm{age}}/
\tau_{\mathrm{age}})}^{\beta_{\mathrm{age}}}\right]$. The aging time
$t_{\mathrm{age}}$ dependence of the relaxation time
$\tau_{\mathrm{age}}$ is defined by these authors through a functional
relation involving the corresponding frequency
$\nu(t_{\mathrm{age}})= 1/(2\pi\tau_{\mathrm{age}})$, but
the stretching exponent $\beta_{\mathrm{age}}$ is same as
the $\beta_\mathrm{\alpha}$, the $\alpha$-relaxation stretching exponent.
We present here an alternative functional form for
$\tau_{\mathrm{age}}(t_{\mathrm{age}})$ directly involving the
relaxation time itself. The proposed model fits the data of
 Lunkenheimer {\em et. al.} perfectly with a stretching exponent
$\beta_\mathrm{age}$ {\em different} from $\beta_\mathrm{\alpha}$.
\end{abstract}

\vspace*{.5cm}

\pacs{64.70.Pf, 77.22Gm, 81.05Kf, 81.40.Tv}

\maketitle
\end{titlepage}

Understanding the dynamics of the supercooled liquid in the non
equilibrium state has been one of the most challenging problems in
condensed matter physics. When cooled fast enough, the supercooled
liquid remains trapped in a specific part of the phase space of the
constituent particles and cannot equilibrate. For such a system the
relaxation time to equilibrium increases far beyond the time scale
of the experiment. Analysis of the dynamics of the liquid in the non
equilibrium state reveals a variety of phenomena like aging
 and memory effects \cite{ediger,struik}.
 Theoretical approaches for understanding the
 complex relaxation behavior in the non equilibrium state include
 computer simulations \cite{kob} and study of simple dynamical models \cite{cugli}.
 An important question in this regard is whether the dynamics in the non equilibrium
 state can be understood as an extrapolation of the alpha relaxation
 process characteristics of  the equilibrium states at temperatures
 above the calorimetric glass transition temperature $T_g$ \cite{ediger}.
 In a recent letter Lunkenheimer
{\em et. al.}\cite{lwsl} have studied the time dependent dielectric
loss data \cite{loidl-de} for various glass formers below the glass
transition temperature. By modeling the experimental data with a
modified Kohlrausch-Williams-Watts (KWW) form,
$\exp\left[-{(t_{\mathrm{age}}/
\tau_{\mathrm{age}}(t_{\mathrm{age}}))}^{\beta_{\mathrm{age}}}\right]$ these authors
demonstrate that the non equilibrium dynamics is fully determined by
the relaxation times and the stretching parameters of the
equilibrium $\alpha$-relaxation process at $T>T_g$. In the present
report we demonstrate that the data of Ref. \cite{lwsl} can also be fitted
in an alternative scheme with a modified KWW form same as that
proposed in Ref. \cite{lwsl}, but with a relaxation time
$\tau_{\mathrm{age}}$ whose time evolution is entirely different.
Furthermore the proposed model which fits the data of Ref.
\cite{lwsl} perfectly obtains a stretching exponent {\em different}
from that of the corresponding $\alpha$-relaxation. It therefore
conforms to a scenario in which the nature of the relaxation  and
heterogeneity controlling the aging process is different from that
of the equilibrium structural relaxation.

First we consider here the data of dielectric loss $\epsilon^"$
\cite{lwsl} for glycerol at 179K ( having calorimetric glass transition
temperature $T_g$ = 185K and fragility $m=53$
\cite{angell,ngai}) over the frequency range 1Hz - $10^5$Hz. During
aging, $\epsilon^"$ for each frequency decreases continuously with
$t_{\rm age}$ approaching the equilibrium value over the longest
time scale of $10^6$ sec. If  the aging data is fitted with the simple stretched
 exponential  form, we obtain stretching
exponent $\beta$ and the relaxation time $\tau$ which are very
different from those corresponding to the equilibrium
$\alpha$-relaxation at $T > T_g$ \cite{sid}.
The data for the dielectric relaxation function is fitted in this
case to the form

\begin{equation}
\label{stret-gen} \epsilon^"\left( t_{\mathrm{age}}\right)  = \left(
\epsilon^"_{\mathrm{st}} - \epsilon^"_{\mathrm{eq}}  \right)
\exp\left[-{(t_{\mathrm{age}}/\tau_{\mathrm{age}})}^{\beta_{\mathrm{age}}}\right]
+ \epsilon^"_{\mathrm{eq}}
\end{equation}

\noindent where the subscripts "st" and "eq" respectively refer to
the initial ($t_{\mathrm{age}} \rightarrow 0$) and the long time
($t_{\mathrm{age}} \rightarrow \infty$) limiting values of
$\epsilon^"$. In such a scheme generally  both
 $\epsilon^"_{\mathrm{st}}$ and $\epsilon^"_{\mathrm{eq}}$
are obtained as the best value for the corresponding fitting
parameters. The relaxation time $\tau_{\mathrm{age}}$ and stretching
exponent $\beta_{\mathrm{age}}$ of the KWW form  are also used as
free fit parameters.  The quality  of this fitting with the dielectric data
from different samples are available in
Ref. \cite{lunken} ( see fig. 1 and 2 there ). Both $\tau_{\mathrm{age}}$ and
$\beta_{\mathrm{age}}$ as obtained from the fitting of the
relaxation data in the aging regime show strong frequency
dependence. If on the other hand we
adopt a fitting scheme in which $\beta_{\mathrm{age}}$ is kept fixed
at its corresponding $\alpha$-relaxation value $\beta_{\alpha} =
.55$, then the data {\em do not} fit the form (\ref{stret-gen}) with
only $\tau_{\mathrm{age}}$ as a free fit parameter. The poorness of
such a fitting procedure is discussed further below (see fig. 4).

An interesting interpretation of this data came from Lunkenheimer
et. al. by using a modified KWW form with the relaxation time
$\tau_{\mathrm{age}}$ being dependent on aging time
$t_{\mathrm{age}}$ \cite{zotev,tool}. However in Ref. \cite{lwsl} the time
dependence of $\tau_{\mathrm{age}}$  is prescribed not in terms of
the relaxation time but the corresponding frequency
$\nu_{\mathrm{age}}$ which is defined as,

\begin{equation}
\label{tau-nu} \nu(t_{\mathrm{age}}) =
\frac{1}{2\pi\tau_{\mathrm{age}}(t_{\mathrm{age}})}.
\end{equation}
\noindent Lunkenheimer et. al. choose the aging time dependence of $\nu$ in (\ref{tau-nu})
in the following form

\begin{equation}
\label{loidl-nu} \nu_{\mathrm{age}} \left( t_{\mathrm{age}} \right)
= \left( \nu_{\mathrm{st}} - \nu_{\mathrm{eq}} \right) \exp \left[
-{\left(2\pi\nu{t_{\mathrm{age}}}\right)}^{\beta_{\alpha}} \right] +
\nu_{\mathrm{eq}}.
\end{equation}

\noindent  According to (\ref{loidl-nu}), the relaxation time
$\tau_{\mathrm{age}}$ $\rightarrow$ $1/(2\pi\nu_{\mathrm{st}})$ and
$1/(2\pi\nu_{\mathrm{eq}})$ as $t_{\mathrm{age}}$ $\rightarrow$ $0$
and $\infty$ respectively. Lunkenheimer et. al. obtain
almost a perfect fit for the dielectric data over the whole frequency
range with the above choice of the time dependence for the relaxation
time. They observe that an important feature of this aging process is that
 the stretching exponent $\beta_{\mathrm{age}}$ is
{\em same} as that of the $\alpha$-relaxation $\beta_{\alpha}$. It
therefore implies that the stretching of the relaxation remains
unaffected by aging below $T_g$ and hence conform to the validity of the time
temperature superposition during the aging process. Furthermore the
best fit value obtained for $\nu_\mathrm{eq}$ in (\ref{loidl-nu})
corresponds to a time $\tau_{\mathrm{eq}} =
1/(2\pi\nu_{\mathrm{eq}})$ which agrees with the $\alpha$-relaxation
time $\tau_\alpha$ for glycerol extrapolated from higher
temperatures ( $T>T_g$ ) to sub-$T_g$ regions. This matching of
$\tau_{\mathrm{eq}}$ with the extrapolated equilibrium $\alpha$-relaxation
 time values in case of glycerol will be further discussed below with fig. 5.


While this is a very instructive way
of interpreting the evolution of the non equilibrium state, the
anasatz (\ref{loidl-nu}) for determining the time evolution of
$\tau_\mathrm{age}$ in terms of the corresponding frequency is
somewhat unnatural. Instead of the relaxation time itself being time
dependent, the corresponding frequency $\nu$ becomes time dependent
in (\ref{loidl-nu}). The latter is then transmitted to
$\tau_{\mathrm{age}}$ through the standard relation (\ref{tau-nu}).
As an alternative to this, in the present paper we adopt a more
natural scheme for the time evolution of the relaxation time
$\tau_{\mathrm{age}}$ in the following manner.

\begin{equation}
\label{our-scal} \tau_{\mathrm{age}}(t_{\mathrm{age}}) = \left(
\tau_{\mathrm{st}}- \tau_{\mathrm{fn}}\right) f(t_{\mathrm{age}}) + \tau_{\mathrm{fn}}~~.
\end{equation}

\noindent where the function $f(t)$ reduces to the value 1 or 0, for
$t \rightarrow 0$  and $\infty$ respectively so that the relaxation
time $\tau_{\mathrm{age}}$ attains the asymptotic values
$\tau_{\mathrm{st}}$ and $\tau_{\mathrm{fn}}$ respectively in the
above two limits. A perfect fit is obtained for all the
$\epsilon^"(t_{\mathrm{age}})$ data at different frequencies using
Eq. (\ref{stret-gen}) with $\tau_{\mathrm{age}}(t_{\mathrm{age}})$
is being determined with the ansatz (\ref{our-scal}). In doing this
the choice of the function $f(t)$ is not unique. However there are
some characteristics that can be associated with this function. As
indicated above it must change from $1$ to $0$ as the time changes
from $t=0$ to $\infty$ limit. It is plausible to expect the time
dependence of $f(t)$ is changing monotonically from 1 to 0. We have
tested in terms of a reduced time $x=t/\tau$, where $\tau$ is the
relaxation time the following three options: (a) exponential ( simple
or stretched) relaxation like $\exp(-x^\beta)$ (b) power law type
decays $x^{-\alpha}$, or (c) a function which decays with the
$\tanh(x)$. Among these the last form gives the best fit to the
aging data. We obtain for the normalized function $f(t)$,

\begin{equation} \label{foft} f(t)= {{a_o} \over
{{[1+\mathrm{e}^{2t/\tau(t)}]}^{\beta}}}
\end{equation}

\noindent where $a_o=2^{1/\beta}$ is a normalization constant to
ensure that $f(t)$ reduces to the above stated values in the two
limits. Note that for times large compared to the relaxation times
this function also {\em behaves like a stretched exponential
function with exponent $\beta$}. Therefore the function defined in
the self-consistent eqn. (\ref{foft}) also represents stretching of
the relaxation process. The exponent $\beta$ for the stretching
process is assumed to be same as the stretching exponent
$\beta_\mathrm{age}$ used in the modified KWW form  to keep the
number of fitting parameters to a minimum. Thus in spite of the
differences in the defining relations for the respective $\tau$'s,
the $\beta_\mathrm{age}$ represents stretching of the relaxation
process in both the cases (Ref.\cite{lwsl} and ours). The fittings
(corresponding to different $\omega$'s) over the whole frequency
range of the dielectric
 data are shown in fig. 1. In the present fitting scheme unlike
that of Ref. \cite{lwsl} we obtain a stretching exponent
 $\beta_\mathrm{age}=.29$ which is {\em different} from the corresponding
 $\alpha$-relaxation value
$\beta_\alpha = .55$. The best fit values for the parameters
$\epsilon^"_{\mathrm{st}}$ and $\epsilon^"_{\mathrm{st}}$ ( at a
given frequency ) obtained respectively in the present work and Ref.
\cite{lwsl}  agree closely. This comparison is displayed in fig 2,
We display next in fig. 3 the aging time dependence of
$\tau_\mathrm{age}$ in our calculation and that of Ref. \cite{lwsl}.
In the present case the the relaxation time is strongly time
dependent at the initial stage and saturates at relatively earlier
as $t_\mathrm{age}$ reach the time $\tau_\mathrm{fn}$ (say). In our
fitting scheme the best fit result value for $\tau_{fn}$ is very
different from the corresponding $\tau_\mathrm{eq}$ of Lunkenheimer
et. al. and cannot be simply related to the $\alpha$ relaxation
process. All the dielectric data for $\epsilon^{"}$ scale into a
single master curve as shown in the fig. 4. In the same figure we
display the curve corresponding to the fitting of Lunkenheimer et.
al. being practically indistinguishable from ours. We also display
here for comparison the  best fittings curves which are obtained
with simple KWW function  having constant relaxation times,
respectively equal to $\tau_{\mathrm{st}}$ and $\tau_{\mathrm{eq}}$.
The stretching parameter  in both cases is $\beta_\alpha$. The
inadequacy of a simple KWW form in fitting the observed data is
clearly observable here.


In addition to the dielectric data for glycerol, following Lunkenheimer
et. al. we also fit using our scheme the dielectric
data for Propylene carbonate ( PC, $T_g \approx 159\mathrm{K}$, $m=104$ \cite{ngai}),
Propylene glycol( PG, $T_g \approx 168\mathrm{K}$, $m=52$ \cite{ngai}), xylitol
 ( $T_g \approx 248\mathrm{K}$, $m=94$ \cite{paluch},\cite{paluch1}) and structural relaxation for
$\mathrm{[Ca(NO_3)_2]_{0.4}[KNO_3]_{0.6}}$ ( CKN, $T_g \approx 333\mathrm{K}$, $m=93$ )
 and the results are respectively shown in fig 5a-5d.
The temperature ( below the corresponding $T_g$ ) in each case is
also displayed in the figure.  The variation of the relaxation time
$\tau_\mathrm{age}(t_\mathrm{age})$ in the modified KWW form for
fitting the data in each of the above materials is shown in fig 6.
For all these materials, the general trend in aging time dependence
of the relaxation time is the same as that in case of glycerol.
$\tau_\mathrm{age}$ is strongly time dependent initially ( up to
time $\tau_\mathrm{fn}$ ) and  finally saturate to an almost
constant value as is clearly visible in fig. 6. The TABLE I
represents the stretching exponent $\beta_\mathrm{age}$ as a best
fit parameter in the aging data for different materials obtained
from present work (column 1) and also from Ref. \cite{lwsl} (column
2).

In case of the fitting scheme of Lunkenheimer et. al. the time scale
$\tau_\mathrm{eq}$ agree with the extrapolated value of the
$\alpha$-relaxation to the sub $T_g$ region. This is shown in fig. 7
for all the five materials whose relaxation data have been
considered above. First $\tau_\alpha$'s for these materials with
inverse temperature $1/T$ are shown. On the same figure we display
for each curve ( with a filled circle ) the corresponding value of
$\tau_{eq} = 1/(2\pi\nu_{eq})$ obtained from the best fit value of
(\ref{loidl-nu}) obtained by Lunkenheimer et. al. These points all
lie on the corresponding extrapolated curve of the
$\alpha$-relaxation times in the sub-$T_g$ regime. These authors
conclude that the equilibrium $\alpha$-relaxation is determining the
 aging process . On the other hand the
corresponding values of $\tau_\mathrm{eq}$ $\equiv$ $\tau_\mathrm{fn}$
obtained in our fitting  scheme are shown for each curve with a filled
triangle conforming to a relaxation mechanism of aging different
from that of the equilibrium $\alpha$-relaxation.

We have presented here an alternative scenario for explaining the
dielectric relaxation data of Lunkenheimer et. al. \cite{loidl-de}
in the aging regime. The present model proposes that the aging
process involve two basic steps. In the first stage, the  aging data
fits to a {\em modified} stretched exponential form with a time
dependent relaxation time $\tau_\mathrm{age}$. This is similar to
the scheme of Lunkenheimer et. al. but the  time dependence of
$\tau_\mathrm{age}(t_\mathrm{age})$  is more natural here. As
$t_\mathrm{age}$ reach a characteristic time scale
$\tau_\mathrm{fn}$ the data  can now be described in this second
stage with a simple KWW form  having a {\em constant} relaxation
time comparable to $\tau_\mathrm{fn}$. This holds simultaneously for
relaxation data at all frequencies as shown in fig 6. The limiting
value of $\tau_\mathrm{fn}$ obtained here from fitting the data is
however {\em not} same as that obtained from extrapolation of
$\alpha$- relaxation times at higher temperatures to the sub-$T_g$
region. The stretching exponent $\beta_\mathrm{age}$ is also very
different from
 that of the $\alpha$-relaxation process. The time dependence of the aging process
and the corresponding relaxation time is possibly controlled by
mechanisms which are different from that of equilibrium
$\alpha$-relaxation.

The  present analysis do not to claim in any way that the fitting
 scheme proposed by Lunkenheimer et. al. is invalid.
The main difference of the model proposed here from that in ref.
\cite{lwsl} lies in the manner in which the relaxation time
$\tau_\mathrm{age}$ depends on $t_{\mathrm{age}}$. In our scheme the
relaxation time $\tau_{\mathrm{age}}$ is directly dependent on
$t_{\mathrm{age}}$. This is distinct from adopting a some what
unusual scheme of (\ref{loidl-nu}) in which the time dependence is
imposed in the frequency $\nu_{\mathrm{age}}$. The justification of
this scheme in which aging time dependence of $\tau_\mathrm{age}$ is
defined in terms of the frequency is not obvious. The motivation for
adopting such a scheme is that the asymptotic value for
$\tau_\mathrm{age}$ (for long times) agrees with value of
$\alpha$-relaxation times at higher temperatures extrapolated to the
sub-$T_g$ region. There is no compelling reason to assume such a
link between the non-equilibrium aging process and the equilibrium
$\alpha$-relaxation.

To explain the last statement  let us consider the correlation function
$\mathcal{C}(t+t_\mathrm{age},t_\mathrm{age})$ of fluctuations at
two different times $t_\mathrm{age}$ and $t+t_\mathrm{age}$. In the
non-equilibrium state $\mathcal{C}$  depends on two times, {\em
i.e.}, both $t$ and $t_\mathrm{age}$. When time translational
invariance holds, the dependence of $\mathcal{C}$ on
$t_\mathrm{age}$ disappears. This happens over some time scale which
is $\tau_\mathrm{eq}$. The correlation
function now depends only on $t$ and the system has reached
equilibrium. At this stage the relaxation is controlled by the
$\alpha$-relaxation time $\tau_\alpha$. But the time scale
$\tau_\mathrm{eq}$ which sets the dependence of the correlation
function on $t_\mathrm{age}$ has no reason to be {\em same} as
$\tau_\alpha$. In other words we insist that there is no aprirori
reason to assume that the  time evolution of $\epsilon^"$ during the
aging process in the non equilibrium state is controlled by the
equilibrium $\alpha$-relaxation process and the corresponding
heterogeneity. Our work clearly shows here that the dielectric data can also
be fitted with the modified KWW form with a different choice for the
aging time dependence of the corresponding relaxation time.

An important aspect of the proposed relaxation behavior lies in the
different time evolution of the relaxation time $\tau$ in comparison
to that of Ref. \cite{lwsl}. Here $\tau$ decreases with waiting time
$t_\mathrm{age}$, implying that the aging process accelerates with
time. It should however be noted that in the present case the
relaxation time only decrease initially and then becomes almost
constant. The stretching data at different frequencies correspond to
a single relaxation time which is almost constant for aging time
$t_\mathrm{age}$ beyond this initial scale $\tau_\mathrm{fn}$. The
stretching exponent $\beta_\mathrm{age}$ for different frequency
data is also the same. In Ref. \cite{lwsl} the relaxation time
$\tau$ actually grows with waiting time. Though somewhat speculative
at this stage, such a difference in the basic nature of the aging
process might imply a different microscopic mechanism for aging
altogether. The more appropriate choice between the two schemes (
for fitting the aging data) discussed here can only be ascertained
with a proper theoretical model for the evolution of the non
equilibrium state.

\vspace*{1cm} SPD gratefully acknowledge A. Loidl and P.
Lunkenheimer for providing the dielectric data and for the
scientific discussions which stimulated the present work. We
 also acknowledge CSIR India for financial support.

\vspace*{1cm}
\begin{center}
\begin{table}[h]
\begin{tabular}[c]{|l|c|l|}
\hline
 &\multicolumn{2}{c|}{Stretching exponent $\beta_\mathrm{age}$}\\[2pt]
\cline{2-3}
Materials & Present Work & \quad Ref. \cite{lwsl} \\[7pt]
\hline
Glycerol & 0.29 & \quad 0.55 \cite{lwsl} \\
PC & 0.24 &\quad0.6 \cite{loidl-de,luk} \\
PG & 0.23 & \quad0.58 \cite{wehn,ngai1} \\
xylitol & 0.45 & \quad0.43 \cite{wehn}  \\
CKN & 0.34 & \quad0.4 \cite{pimenov} \\
\hline
\end{tabular}
\caption{The stretching exponent $\beta_\mathrm{age}$ as a best fit
parameter in the aging data for different materials obtained (a)
from present work (column 1) (b) from Ref. \cite{lwsl} (column 2).
The corresponding reference from which the data for
$\alpha$-relaxation is taken are also provided in col. 2}
\end{table}
\end{center}

\newpage

\newpage
\begin{figure}
\includegraphics*[width=8cm]{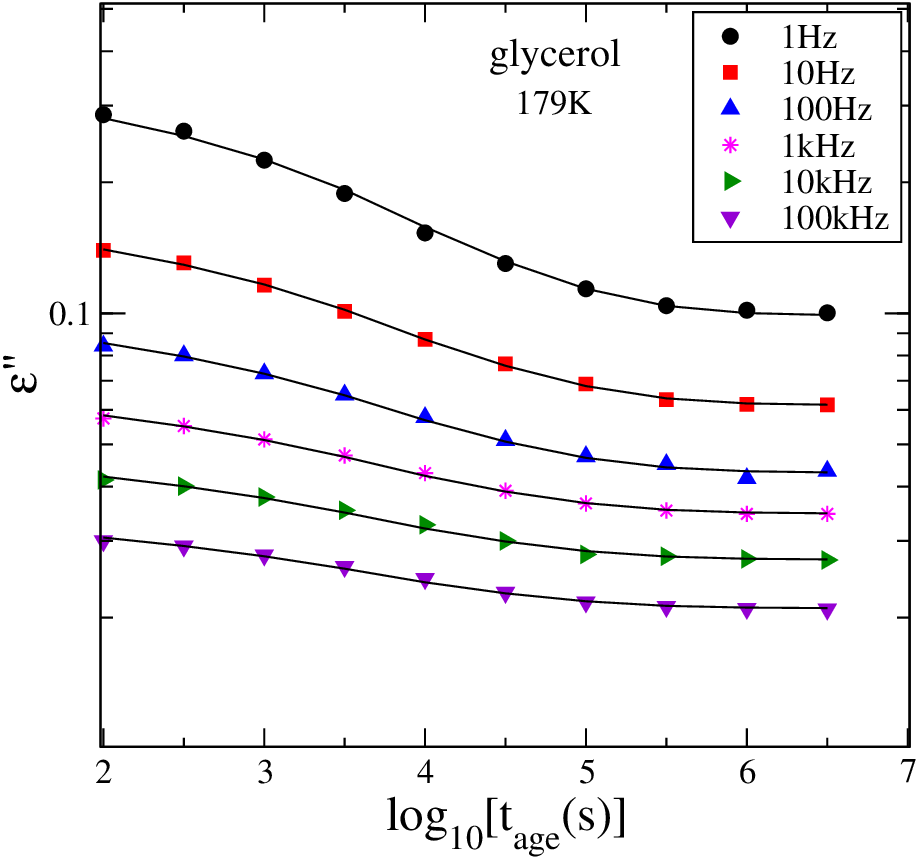}
\caption{ $\epsilon^"(t_{\mathrm{age}})$ data for glycerol at $179^oK$ vs.
$t_{\mathrm{age}}$ (in sec) at the frequencies displayed in the
inset. Solid lines correspond to the best fits using the modified
KWW form with $\tau_{\mathrm{age}}(t_{\mathrm{age}})$ determined
from eqn. (\ref{our-scal}).}
\end{figure}

\begin{figure}
\includegraphics*[width=8cm]{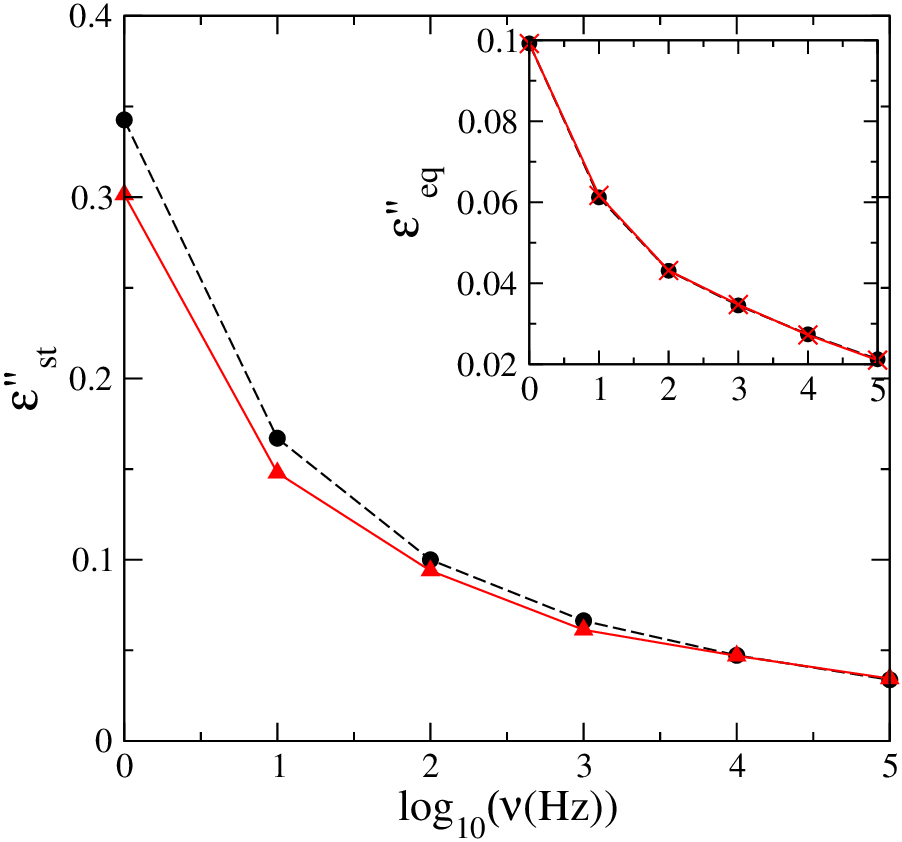}
\caption{ The best fit values of
$\epsilon^"_{\mathrm{st}}$ for the curves in fig. 1
vs. frequency $\nu$(Hz) as obtained with the present work (solid line)
and the corresponding results of Ref. \cite{lwsl} (dashed). The inset shows the
best fit $\epsilon^"_{\mathrm{eq}}$ with frequency $\nu$ respectively from
the present work and Ref. \cite{lwsl}.}
\end{figure}

\newpage
\begin{figure}
\includegraphics*[width=8cm]{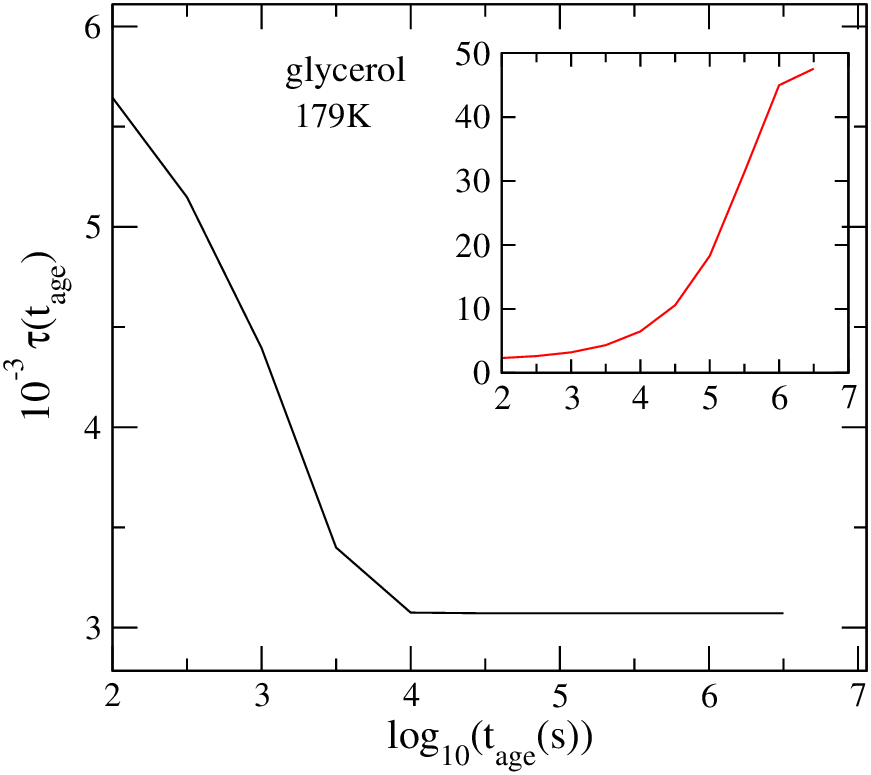}
\caption{ The best fit $\tau(t_{\mathrm{age}})$  vs. $t_{\mathrm{age}}$
 (in sec) for glycerol using ansatz
 (\ref{our-scal}) of the present work.
 Result of Ref. \cite{lwsl} for $\tau(t_{\mathrm{age}})$ as inset.}
\end{figure}

\begin{figure}
\includegraphics*[width=8cm]{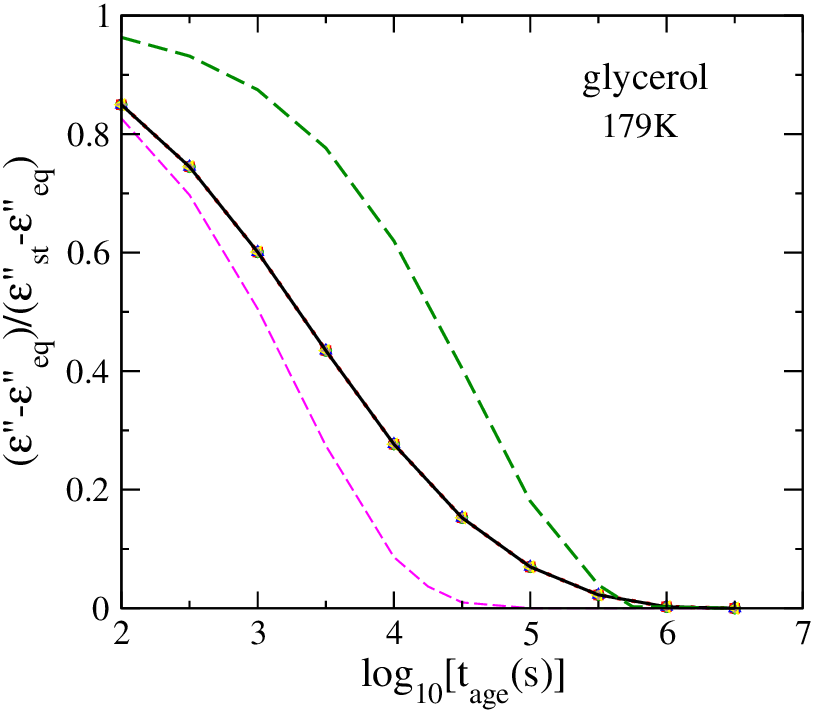}
\caption{ Scaling of the dielectric data for glycerol
at $179^oK$ corresponding to different frequencies.
Fits obtained with $\tau_{\mathrm{age}}(t_{\mathrm{age}})$ in the
modified KWW ansatz determined from
(a) Eqn. (\ref{our-scal}) of the present work (solid line) and (b)
from Ref. \cite{lwsl} (dashed line). Also shown are best fits
obtained taking $\tau_{\mathrm{age}}$
= $\tau_{\mathrm{st}}$ (short dashed) and $\tau_{\mathrm{age}}$ =
$\tau_{\mathrm{eq}}$ (long dashed) respectively with the
 stretching exponent being fixed at $\beta_\alpha$.}
\end{figure}

\newpage
\renewcommand{\thefigure}{\arabic{figure}\alph{subfigure}}
\setcounter{subfigure}{1}
\begin{figure}
\includegraphics*[width=8cm]{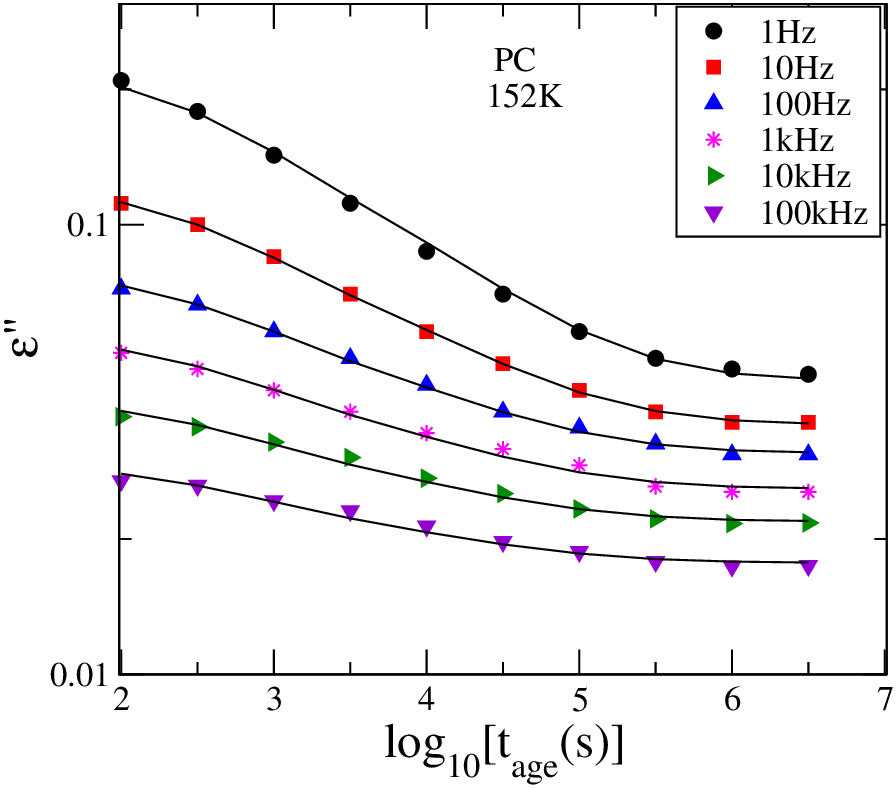}
\caption{ $\epsilon^"(t_{\mathrm{age}})$ data for PC \cite{loidl-de,luk}
at $152^oK$ vs. $t_{\mathrm{age}}$ (in sec) at the frequencies displayed in the
inset. The solid lines are obtained using our scheme.}
\end{figure}

\addtocounter{figure}{-1}
\addtocounter{subfigure}{1}
\begin{figure}
\includegraphics*[width=8cm]{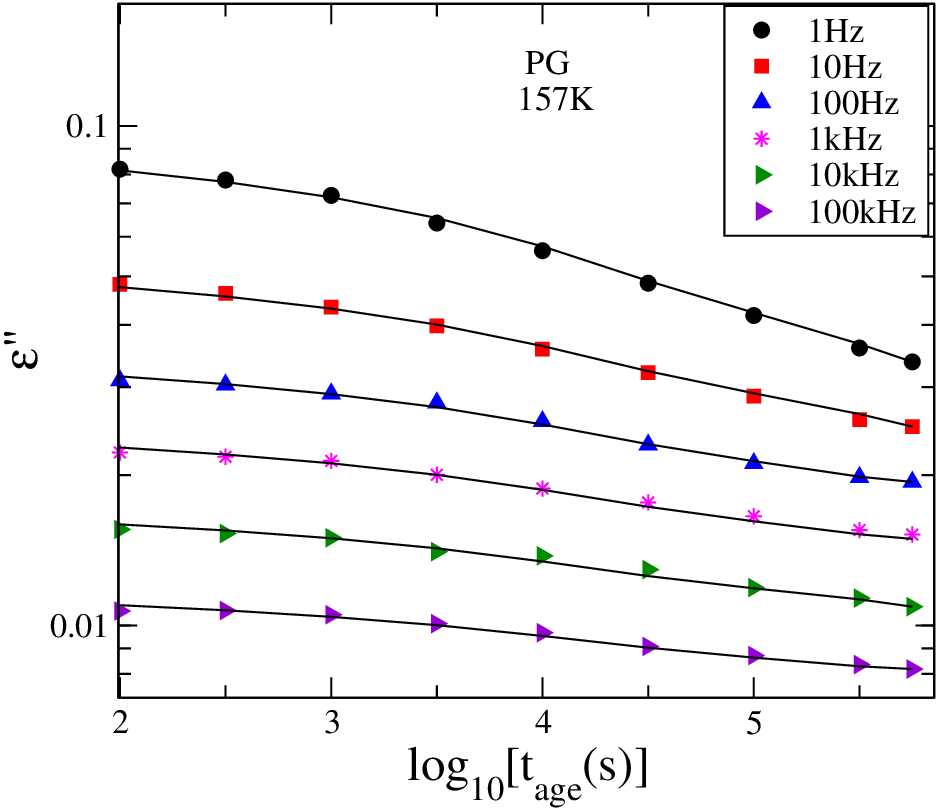}
\caption{ $\epsilon^"(t_{\mathrm{age}})$ data for PG \cite{wehn,ngai}
at $157^oK$ vs. $t_{\mathrm{age}}$ (in sec) at the frequencies displayed in the
inset. The solid lines are obtained using our scheme.}
\end{figure}

\newpage
\addtocounter{figure}{-1}
\addtocounter{subfigure}{1}
\begin{figure}
\includegraphics*[width=8cm]{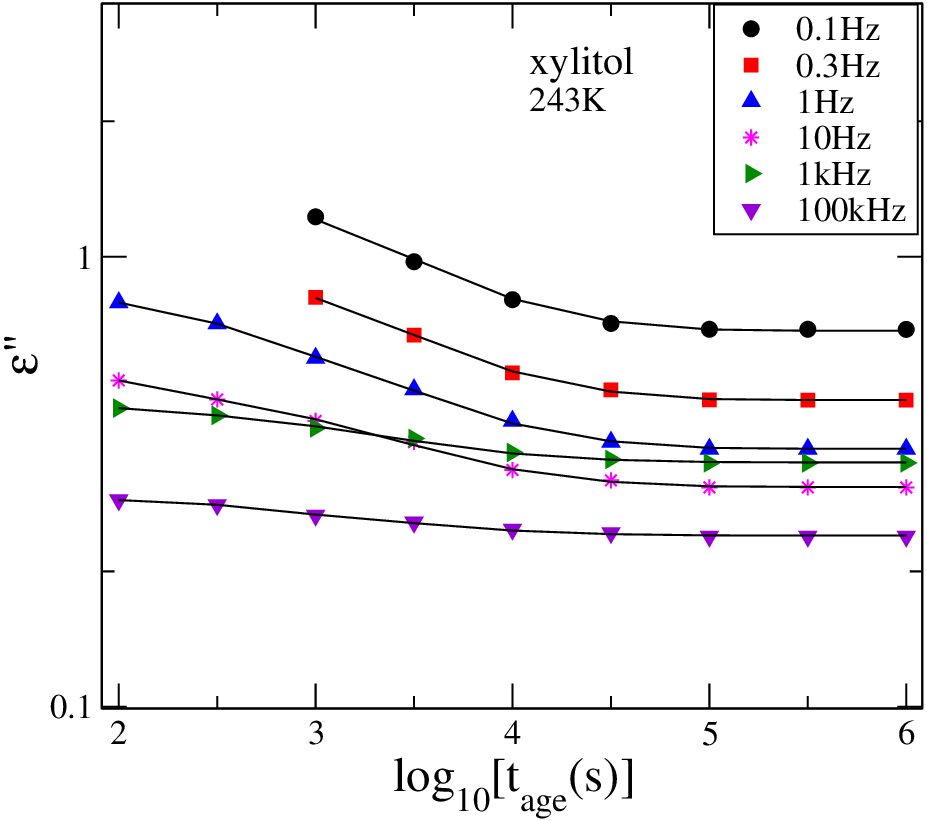}
\caption{ $\epsilon^"(t_{\mathrm{age}})$ data for xylitol \cite{wehn}
at $243^oK$ vs. $t_{\mathrm{age}}$ (in sec) at the frequencies displayed in the
inset. The solid lines are obtained using our scheme.}
\end{figure}

\addtocounter{figure}{-1}
\addtocounter{subfigure}{1}
\begin{figure}
\includegraphics*[width=8cm]{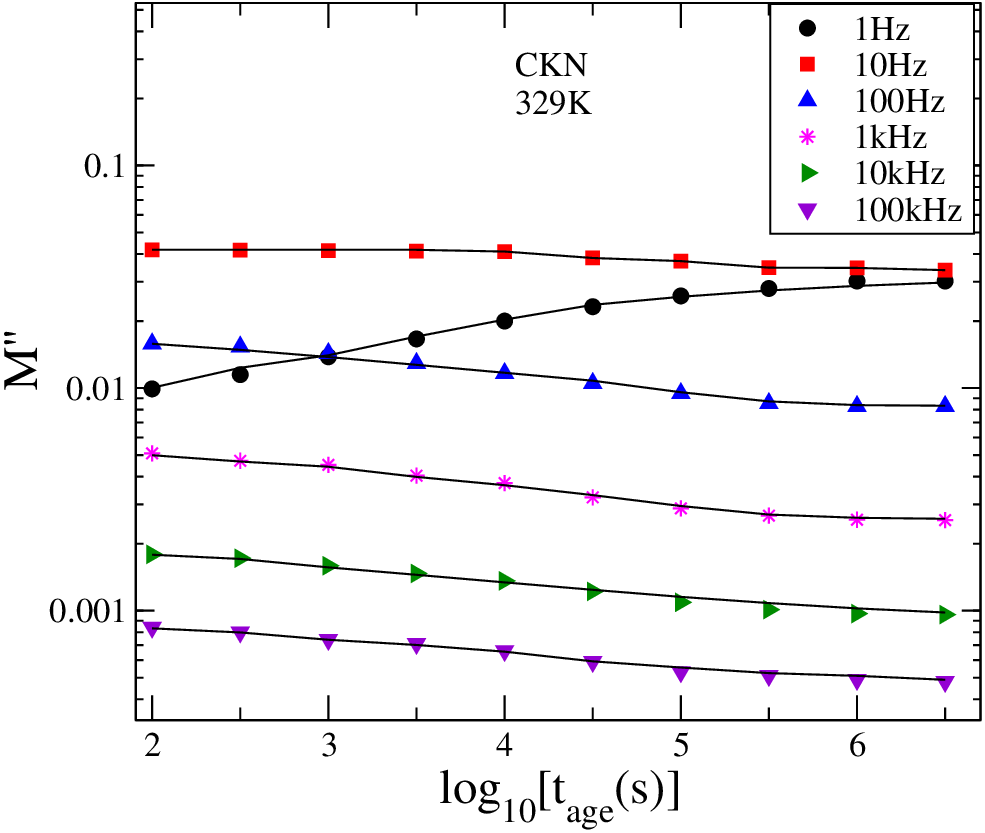}
\caption{ $M^"(t_{\mathrm{age}})$ data for CKN \cite{pimenov}
at $329^oK$ vs. $t_{\mathrm{age}}$ (in sec) at the frequencies displayed in the
inset. The solid lines are obtained using our scheme.}
\end{figure}

\newpage
\renewcommand{\thefigure}{\arabic{figure}}
\begin{figure}
\includegraphics*[width=8cm]{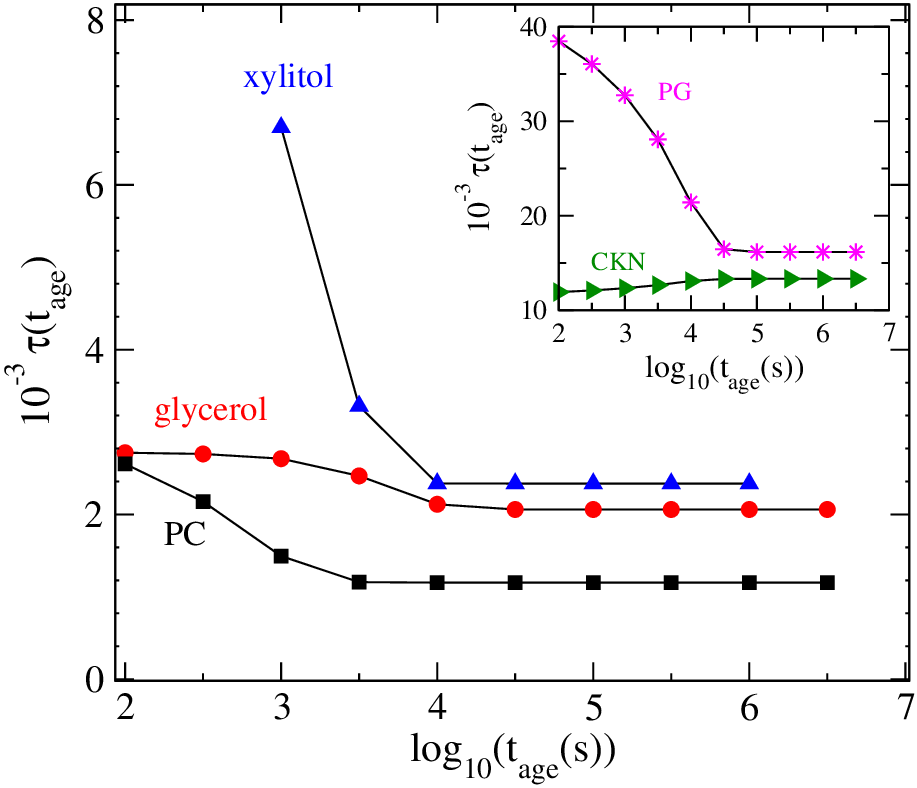}
\caption{ $\tau(t_{\mathrm{age}})$  vs. $t_{\mathrm{age}}$ (in sec) for different materials using ansatz (\ref{our-scal}) in the
present work.}
\end{figure}

\begin{figure}
\includegraphics*[width=8cm]{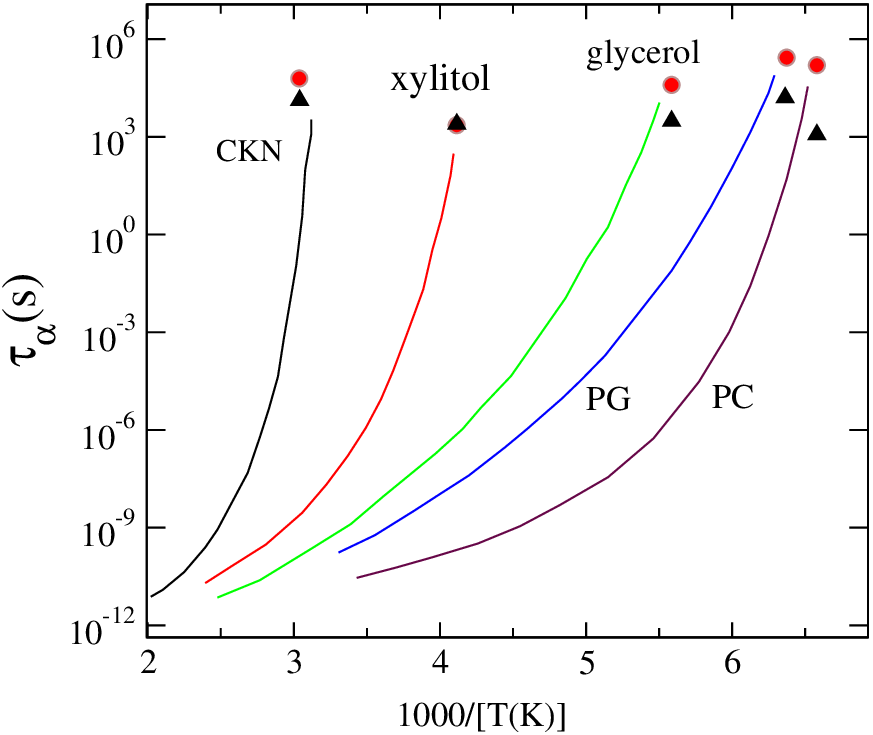}
\caption{ $\alpha$-relaxation time
$\tau_\alpha$ vs. inverse of temperature for CKN\cite{loidl-de}
, xylitol\cite{luk},
glycero\cite{wehn},
PG\cite{ngai1},
 PC\cite{pimenov}.
 The  $\tau_\mathrm{eq}$ obtained from the fitting of the corresponding
 data is shown by filled circles for the fitting of Ref. \cite{lwsl}
 and by filled triangles the present fitting scheme. }
\end{figure}


\vspace*{1cm}
\begin{thebibliography}{99}


\bibitem{ediger}
M.D. Ediger, C.A. Angell, and S.R. Nagel, J. Phys.
Chem. {\bf 100}, 13200(1996); C.A. Angell {\em et al.}, J. Appl.
Phys. {\bf 88}, 3113 (2000)

\bibitem {struik}
L.C.E. Struik, {\em Physical Aging In Amorphous Polymers and Other Materials}
 (Elsevier, Amsterdam, 1978)

\bibitem{kob}
W. Kob and J. L. Barrat, Phys. Rev. Lett. {\bf 38}, 4581 (1997).

\bibitem{cugli}
L.F. Cugliandolo and J. Kurchan, Phys. Rev. Lett. {\bf 71}, 173 (1993).

\bibitem {lwsl}
P. Lunkenheimer, R. Wehn, U. Schneider, and A. Loidl, Phys. Rev.
Lett. {\bf 95}, 055702 (2005).


\bibitem {loidl-de}
U. Schneider, R. Brand, P. Lunkenheimer, and A. Loidl, Phys. Rev.
Lett. {\bf 84}, 5560 (2000).

\bibitem {angell}
C.A. Angell, in {\em Relaxations in Complex Systems}, edited by
K.L. Ngai and G.B. Wright (NRL, Washington, DC,1985), p. 3.

\bibitem{ngai}
R. B\"{o}hmer, K.L. Ngai, C.A. Angell, and D.J. Plazek, J.Chem. Phys.
{\bf 99}, 4201 (1993)

\bibitem {sid} R. L. Leheny and S. R. Nagel, Phys. Rev. B {\bf 57}, 5154 (1998).

\bibitem{lunken}
P. Lunkenheimer, R. Wehn, A. Loidl, J. Non-Cryst. Solids
{\bf 352} 4941(2006)

\bibitem{zotev} V. S. Zotev, G. F. Rodriguez, G. G. Kenning, R. Orbach, E. Vincent and J. Hammann, Phys. Rev. B {\bf 67}, 184422 (2003)

\bibitem {tool} A.Q. Tool, J.Am. Ceram. Soc. {\bf 29}, 240 (1946); O.S. Narayanaswamy, {\em ibid}. {\bf 54}, 240 (1971).

\bibitem{paluch}
A. D\"{o}\ss, M. Paluch, H. Sillescu, and G. Hinze, Phys. Rev.
Lett. \textbf{88}, 095701 (2002).

\bibitem{paluch1} A. D\"{o}\ss, M. Paluch, H. Sillescu, and G. Hinze, J. Chem. Phys. \textbf{117}, 6582 (2002); A. Minuguchi, K. Kitai, and R.
Nozaki, Phys. Rev. E \textbf{68}, 031501 (2003).

\bibitem{luk}  P. Lunkenheimer, U. Schneider, R. Brand, and A. Loidl,
Contemp. Phys. {\bf 41} 15 (2000).

\bibitem {wehn} R. Wehn {\em et al}. (to be published).

\bibitem{ngai1} K. L. Ngai et al., J. Chem. Phys. {\bf 115}, 1405 (2001).

\bibitem{pimenov} A. Pimenov, P. Lunkenheimer, H. Rall, R. Kohlhaas, A. Loidl and R B\"{o}hmer, Phys. Rev. E  {\bf 54}, 676 (1996).


\end{thebibliography}
\end{document}